\documentclass[aps,twocolumn]{revtex4}
\usepackage{epsfig}
\usepackage{psfrag}
\newcommand{\be}{\begin{equation}}
\newcommand{\ee}{\end{equation}}
\newcommand{\ba}{\begin{eqnarray}}
\newcommand{\ea}{\end{eqnarray}}

\begin{document}

\title{Polymers grafted to porous membranes}

\author{Hsiao-Ping Hsu$^{a,b}$ and Peter Grassberger$^{a}$}

\affiliation{$^a$John-von-Neumann Institute for Computing, Forschungszentrum J\"ulich,
D-52425 J\"ulich, Germany\\
and\\
$^b$Institute of Physics, Johannes Gutenberg University of Mainz\\
D-55099 Mainz, Staudinger Weg 7, Germany}

\date{\today}

\begin{abstract}
We study a single flexible chain molecule grafted to a membrane which has pores
of size slightly larger than the monomer size. On both sides of the membrane 
there is the same solvent. When this solvent is good, i.e. when the polymer
is described by a self avoiding walk, it can fairly easily penetrate the 
membrane, so that the average number of membrane crossings tends, for chain 
length $N\to\infty$, to a positive constant. The average numbers of monomers 
on either side of the membrane diverges in this limit, although their ratio 
becomes infinite. For a poor solvent, in contrast, the entire polymer is located,
for large $N$, on one side of the membrane. For good and for theta solvents 
(ideal polymers) we find scaling laws, whose exponents can in the latter case 
be easily understood from the behaviour of random walks.
\end{abstract}

\maketitle
%\section{Introduction}

Long chain polymers in dilute solutions have important technical applications,
but they also form an interesting model for studying scaling and renormalization 
group techniques \cite{deGennes,Grosberg,Schafer}. As a consequence, most of 
their properties have been studied in detail and are well understood, at least 
as far as equilibrium is concerned. In this paper we want to study a new effect 
which, to our knowledge, has not been considered in the previous literature, 
although also it has a number of possible applications.

Physically, we ask ourself what is the influence of the solvent on the ability
of a polymer to penetrate a porous membrane. If the pores are very big, 
typically of sizes larger than a collapsed globule but smaller than the 
Flory diameter of a random coil in a good solvent, the collapse might help
in penetrating them. On the other hand, we expect the opposite effect for 
small pores, typically just slightly larger than a single monomer. In that 
case a polymer in good solvent can go through a pore by having one monomer 
pass after the other \cite{barkema}. For a collapsed polymer this should be 
less easy, since monomers tend to cluster together. But, as far as we know, 
no detailed studies of this exist.

Technically, we consider self avoiding walks (SAWs) of $N$ steps on a simple cubic lattice. In
a good solvent at temperature $T\to\infty$, the only monomer-monomer interaction 
is the excluded volume interaction. In a less good solvent, we include an 
attractive monomer-monomer contact potential which acts between unbonded 
monomer pairs on nearest neighbour sites on the lattice, and has 
strength $-\epsilon$, giving for each such contact a Boltzmann factor 
$q = e^{\epsilon/K_bT}$. The theta point for this model is at $q_\theta = 1.3087\pm 0.0003$
\cite{PERM}. The membrane is located in the plane $z=0$. Its pores are of 
unit size, and are arranged in checkerboard fashion: sites with even $x+y$ and 
$z=0$ are forbidden for the walk, while all other sites are allowed. The walks
start at the origin $x=y=z=0$, i.e. one end is grafted to one of the forbidden
sites of the membrane.

We simulate this model using the pruned-enriched Rosenbluth method (PERM) \cite{PERM}.
This is the most efficient known method for simulating theta polymers, but it is 
also very efficient for conditions not too far from the theta point, including
athermal SAWs. For the latter, the pivot algorithm is even more efficient if the
SAWs are not constraint by obstacles like our membrane. If there are many such 
obstacles, PERM becomes more efficient. We have not checked whether the pivot 
method would be more efficient for the present problem, but this is of little 
relevance since we were anyhow able to obtain very high statistics data for very long 
chains ($N=16000$ at the theta point, $N=10000$ for athermal SAWs).

To measure how well the polymer can penetrate through the membrane, we use two 
different ``order parameters" (i.e., observables). The first is simply the number 
$N_0$ of points on the walk in the plane $z=0$ (excluding the starting point). For the 
other, we first define $N_+$ as the number of monomers in the upper half space $z>0$, 
and $N_-$ as the number of monomers at $z<0$. Obviously, $N_++N_-+N_0 = N$. The 
second order parameter is then
\be 
   \alpha = {2 N_+ N_- \over N_+^2+N_-^2}\;.
\ee

% Fig. 1
\begin{figure}
  \begin{center}
\psfig{file=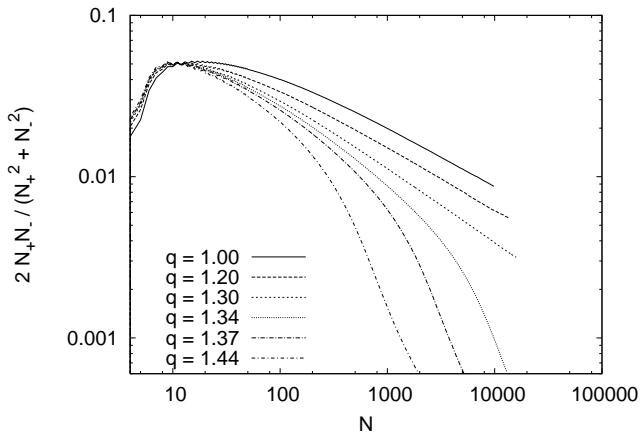,width=5.9cm, angle=270}
   \caption{Log-log plot of $\alpha$ against $N$, for six different values of $q$.
     Data decrease with $q$.
     Statistical errors are not larger than the thickness of the lines.}
   \label{1}
  \end{center}
\end{figure}

% Fig. 2
\begin{figure}
  \begin{center}
\psfig{file=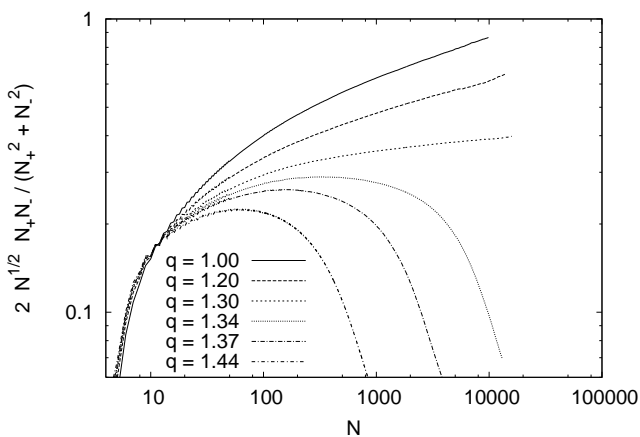,width=5.9cm, angle=270}
   \caption{Same data as in Fig.~1, but multiplied by $N^{1/2}$.}
   \label{2}
  \end{center}
\end{figure}

In Fig.~1 we plot $\alpha$ against $N$, for six different values of $q$. Apart from 
the strong finite size effects at small $N$ ($\alpha$ vanishes trivially for $N<4$),
there seems to be a power behaviour for $q \leq q_\theta$. Indeed, when plotting
$\alpha\sqrt{N}$ against $N$ (Fig.~2), we see that  
\be
   \alpha \sim N^{-1/2}     \label{alpha-theta}
\ee
for theta polymers, while $\alpha$ seems to decrease with a smaller power for 
polymers in a good solvent. Due to the aforementioned scaling corrections it is 
hard to give a more precise estimate, our best fit is $\alpha \sim N^{-0.38}$
for SAWs. For collapsed chains $\alpha$ decreases much faster, but it is not 
clear whether it will finally follow a power law (with power $\geq 1$) or whether 
it falls faster.

% Fig. 3
\begin{figure}
  \begin{center}
\psfig{file=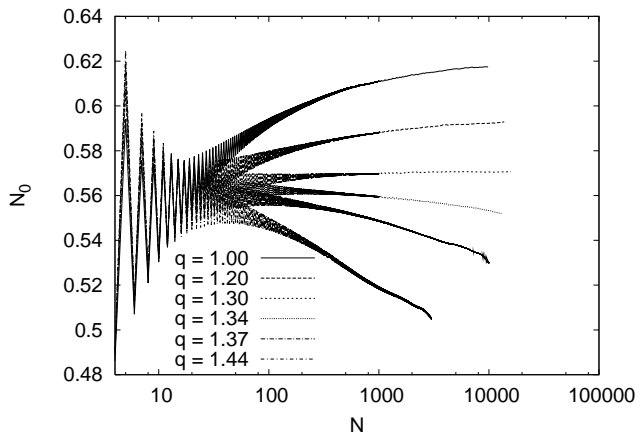,width=5.9cm, angle=270}
   \caption{Log-linear plot of $\langle N_0\rangle$ against $N$, for the same
    six different values of $q$. Again, data decrease with $q$. The rapid 
    decrease of the data for $q=1.37$ and $q=1.44$ at very large $N$ is 
    presumably due to statistical fluctuations.}
   \label{3}
  \end{center}
\end{figure}

The average $N_0$ is plotted against $N$, for the same six values of $q$, in 
Fig.~3. We see huge even/odd oscillations for small $N$. Apart from that, the results 
are somewhat complementary to those in Figs.~1 and 2. For $q=q_\theta$ we see that 
$N_0$ converges quickly to a constant, 
\be
    N_0 \to 0.5703\pm 0.0003 \qquad {\rm for} \quad N\to\infty \;.
\ee
Again scaling corrections are more important for good solvent conditions. But although 
$N_0$ keeps increasing with $N$ in this case, it seems to converge nevertheless to 
a constant. This is definitely no longer true for $q>q_\theta$. There, $N_0$ seems 
to converge to zero. We cannot exclude that this happens according to power laws, 
$N_0 \sim N^{-\beta}$, but then the exponent $\beta$ would be non-universal, and 
would increase strongly with $q$. We consider it much more likely that $N_0$ 
decreases, in the collapsed phase, faster than any power of $N$.

Let us now try to understand these results theoretically, and let us start with 
$q=q_\theta$. For simplicity we will neglect the logarithmic corrections holding 
at the theta point \cite{duplantier,hegger,PERM,hager}, and describe the polymers by 
ideal random walks. In this case, the walks are Markovian. Whenever such a walk
arrives at $z=0$, it will go in the next step to $z>0$ and $z<0$ with equal
probability, and both possibilities will evolve later in the same way (except for 
mirror symmetry). Thus, as far as scaling is concerned, this is the same situation
as for an impenetrable wall. In this case \cite{eisenriegler} the monomer density 
decreases near the wall as $z^2$, and the number of monomers just touching the wall
goes to a constant for $N\to\infty$. This is in agreement with Fig.~3. 

The probability
that the $n$-th monomer, with $0\ll n\ll N$, is at distance $z$ from the wall scales 
more precisely as $z^2/n^{3/2}$, where the $n$-dependence is forced by the fact 
that $\langle z\rangle \sim \sqrt{n}$. Thus the probability that the wall is touched 
by the $n$-th monomer scales as $n^{-3/2}$, and this will also be the scaling of 
the probability that the walk passes through the membrane at ``time" $n$.
From this and the Markov property follows immediately Eq.(\ref{alpha-theta}).

These simple arguments break down when $q\neq q_\theta$. Now, it does matter that 
a walk which passes through a pore ends up on the opposite side of the membrane.
For $q<q_\theta$ (good solvent conditions) it feels then a weaker excluded volume
effect, as compared to the case where it stayed on the same side of the membrane.
Thus passing through the membrane is (entropically) favoured, leading to larger 
values of $N_0$ and $\alpha$. On the other hand, passing through the membrane is 
(energetically) disfavoured for bad solvent conditions, and $N_0$ and $\alpha$
are reduced in this case. Apart from this very qualitative argument we have not 
been able to give a more thorough theoretical discussion. For SAWs grafted to a
hard wall, the average number of monomers touching the wall goes to a constant for 
$N\to\infty$ \cite{eisenriegler,hsu}. Thus it is not too surprising that the same 
holds also when the wall is replaced by a porous membrane, and only monomers 
within pores are counted. But it is not too obvious 
either, as the same analogy does not hold for collapsed polymers. While we found 
here that $N_0$ goes to zero for $N\to\infty$, the number $N_1$ of monomers {\it touching}
the wall or membrane ($z=\pm 1$; here it does not matter whether the membrane has 
pores or not) diverges with $N$, when the solvent is bad ($q>q_\theta$; see 
Fig.~4).

% Fig. 4
\begin{figure}
  \begin{center}
\psfig{file=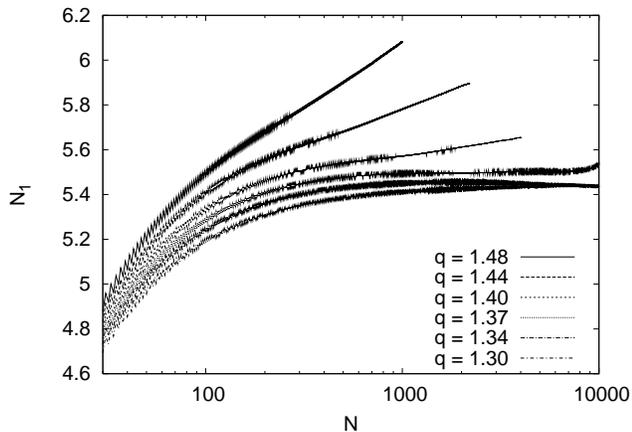,width=5.9cm, angle=270}
   \caption{Number of monomers at distance $z=\pm 1$ from the membrane, $N_1$, 
    plotted against $\ln N$ for several values of $q$. While $N_1$ 
    converges to a constant when $q\leq q_\theta$, it seems to diverge with $N$ 
    for $q>q_\theta$.}
   \label{4}
  \end{center}
\end{figure}

% Fig. 5
\begin{figure}
  \begin{center}
\psfig{file=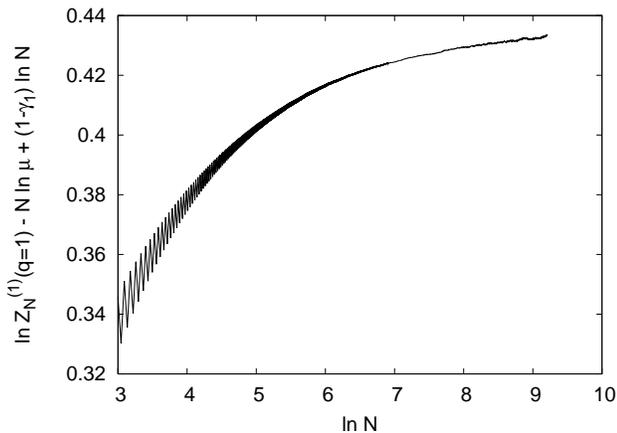,width=5.9cm, angle=270}
   \caption{Logarithm of the partition sum divided by $\mu^N$ and multiplied by 
    $N^{1-\gamma_1}$, plotted against $\ln N$. The values used for $\mu$ and $\gamma_1$
    are $4.684038$ and $0.679$. They were fixed by demanding that the leading 
    corrections to scaling go like $N^{-1/2}$.}
   \label{5}
  \end{center}
\end{figure}

% Fig. 6
\begin{figure}
  \begin{center}
\psfig{file=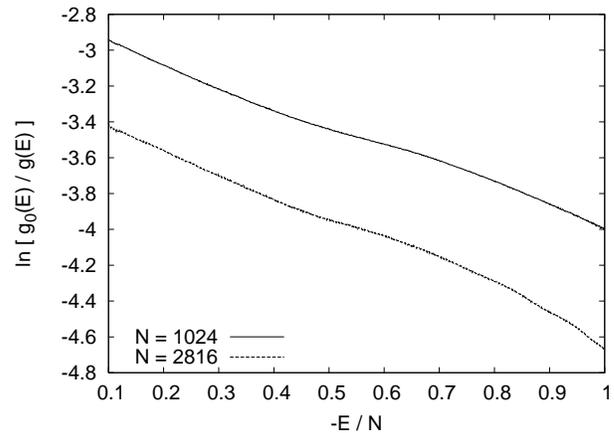,width=5.9cm, angle=270}
   \caption{Ratio between the density of states for polymers grafted to a porous
    membrane, divided by the analogous density of states for free polymers without
    a membrane. The logarithm of this ratio is plotted versus $-E/N$, for two 
    different values of $N$.}
   \label{6}
  \end{center}
\end{figure}

Finally, let us discuss how the porous membrane effects the partition function and 
the density of states $g(E)$, where $E = -\epsilon m$ with $m$ being the number of 
monomer-monomer contacts. For SAWs grafted to an impenetrable wall, the partition 
sum scales as \cite{gr-hegg}
\be
   Z_N^{(1)} \sim \mu^N N^{\gamma_1-1}
\ee
where $\mu = 4.684043(10)$ is the same effective connectivity as for ordinary SAWs, 
but $\gamma_1$ is a modified critical exponent, $\gamma_1 = 0.679(2)$. Numerically
(see Fig.~5) we found that the same ansatz describes also the large $N$ behaviour 
of athermal SAWs grafted to the porous membrane (in Fig.~5 we used our present best 
estimates $\gamma_1 = 0.679(1)$ and $\mu = 4.684038(6)$). This is not trivial, since 
we had just argued that the possibility of penetrating the membrane should increase 
the partition sum. It does so indeed, but this does not affect the asymptotic 
behaviour.

Instead of showing the density of states $g(E)$ itself, defined as the number 
of configurations with energy $-E$, we show in Fig.~6 the ratio $g(E)/g_0(E)$, 
where $g_0(E)$ refers to polymers in absence of a membrane. More precisely, we 
plot the logarithm of this ratio against $-E/N$ for two values of $N$. Here we have 
assumed that $\epsilon=1$, i.e. each contact contributes exactly one unit of energy.
We see that $g(E)/g_0(E)$ decreases with the absolute value 
of $E$, i.e. the effect of the membrane increases with $|E|$. This is of course
in agreement with our other measurements. A more careful inspection shows 
that the curves in Fig.~6 are not straight, but are cup-concave around $E \approx 
-0.4 N$. This means that $g(E)$ is less convex than $g_0(E)$ in the energy range 
which gives the dominant contributions to the partition sum near the theta 
collapse, i.e. the theta transition becomes ``harder"  by the presence of the 
membrane. For the present model this is a rather weak effect, but is is not 
inconceivable that it is much stronger in some other microscopic realization. In 
such a model, grafting polymers to porous membranes may make the collapse 
transition look first order for intermediate values of $N$.

We thank Walter Nadler for very helpful discussions.

\end{document}